\documentclass[letterpaper, 10 pt, conference]{ieeeconf}  

\IEEEoverridecommandlockouts                              

\usepackage {booktabs}
\usepackage{url}
\usepackage{hyperref}
\usepackage{multirow}

\usepackage{cite}
\usepackage{amsmath,amssymb,amsfonts}
\usepackage{algorithmic}
\usepackage{graphicx}
\usepackage{textcomp}
\usepackage{xcolor}

\def\BibTeX{{\rm B\kern-.05em{\sc i\kern-.025em b}\kern-.08em
    T\kern-.1667em\lower.7ex\hbox{E}\kern-.125emX}}

\title{\LARGE \bf
Assessing the Utility of Audio Foundation Models for Heart and Respiratory Sound Analysis
}

\author{Daisuke~Niizumi, Daiki~Takeuchi, Masahiro Yasuda, Binh Thien Nguyen,  \\
Yasunori~Ohishi,~\IEEEmembership{Member,~IEEE,} and~Noboru~Harada,~\IEEEmembership{Senior~Member,~IEEE}
\thanks{The authors are with NTT Communication Science Laboratories, Nippon Telegraph and Telephone Corporation, Atsugi 243-0198, Japan
(e-mail: daisuke.niizumi@ntt.com; d.takeuchi@ntt.com; masahiro.yasuda@ntt.com; binhthien.nguyen@ntt.com; yasunori.ohishi @ntt.com; harada.noboru@ntt.com)}
}

\begin{document}

\maketitle
\thispagestyle{empty}
\pagestyle{empty}

\begin{abstract}
Pre-trained deep learning models, known as foundation models, have become essential building blocks in machine learning domains such as natural language processing and image domains. This trend has extended to respiratory and heart sound models, which have demonstrated effectiveness as off-the-shelf feature extractors. However, their evaluation benchmarking has been limited, resulting in incompatibility with state-of-the-art (SOTA) performance, thus hindering proof of their effectiveness.
This study investigates the practical effectiveness of off-the-shelf audio foundation models by comparing their performance across four respiratory and heart sound tasks with SOTA fine-tuning results. Experiments show that models struggled on two tasks with noisy data but achieved SOTA performance on the other tasks with clean data. Moreover, general-purpose audio models outperformed a respiratory sound model, highlighting their broader applicability.
With gained insights and the released code, we contribute to future research on developing and leveraging foundation models for respiratory and heart sounds.
\newline
\indent \textit{Clinical relevance}— This study reveals the practical effectiveness of the publicly available audio foundation models for respiratory and heart sound diagnosis applications.
\end{abstract}

\section{Introduction}
Foundation models\cite{Bommasani2021FoundationModels}, deep learning models pre-trained on a large-scale dataset, have advanced and become essential components in AI systems for various applications. Following this trend, there is growing interest in audio foundation models for respiratory and heart sounds\cite{OPERA,baur2024HeAR,FM_cardiovascular}, with a focus on exploring their potential as universal feature extractors for analyzing these sounds captured through stethoscopes or microphones.

Although these methods showcase the efficacy of the proposed foundation models within their respective evaluation benchmarks, these benchmarks fall short of offering comparisons with state-of-the-art (SOTA) performance. Even though these evaluations are conducted on a variety of open datasets\cite{OPERA}, they lack alignment with commonly used evaluation procedures and metrics, such as those employed in the respiratory sound task ICBHI2017\cite{ICBHI2017}. Furthermore, only one of these models has been made public\cite{OPERA}, even though they have been proposed as foundation models.
Consequently, the practical effectiveness of the proposed models remains unverified.

On the other hand, we have demonstrated that foundation models can reach SOTA performance when fine-tuned\cite{Exploring}. However, with advances in large language model (LLM) technology, the demand for high-performance, off-the-shelf models with fixed weights is rising to address the growing interest in developing innovative multimodal LLM applications that leverage natural language queries\cite{MLLM,RespLLM}. In this context, audio foundation models are essential building blocks as fixed-weight audio feature extractors.

This study aims to answer the question about the performance of foundation models compared to SOTA results without fine-tuning.
To do this, we evaluate the models with fixed weights and compare their performance with SOTA results under commonly used evaluation procedures and metrics for the tasks.
Experiments show that recent general-purpose audio foundation models pre-trained on a large-scale general audio dataset (e.g., environmental sound, speech, or music) attain SOTA or baseline performance for both respiratory and heart sounds, but that one pre-trained on respiratory sounds underperforms in respiratory tasks.
With the results and insights gained in the experiments, this study helps the further development and effective use of foundational models in future AI systems for respiratory and heart sounds. Our evaluation code is available online.\footnote{\url{https://github.com/nttcslab/eval-audio-repr/tree/main/app}}

\section{Experiments}
We assessed the effectiveness of the audio foundation models with fixed weights in experiments involving two respiratory sound tasks (Sections \ref{sec:exp-ICBHI} and \ref{sec:exp-SPRS}) and two heart sound tasks (Sections \ref{sec:exp-CirCor} and \ref{sec:exp-BMD-HS}) by comparing their performance to SOTA and baseline models with fine-tuning.

\subsection{Evaluated Audio Foundation Models} \label{sec:exp-models}
We evaluated the audio foundation models AST\cite{gong2021ast}, BYOL-A\cite{niizumi2023byol-a}, and M2D\cite{M2D2024TASLP}, which have been evaluated with fine-tuning for heart sounds in \cite{Exploring}; BEATs\cite{chen2022beats}, which has been utilized in many audio applications studies; and OPERA-CT\cite{OPERA}, a respiratory sound foundation model pre-trained specifically on respiratory sounds. No heart sound foundation model is publicly available.
AST, BYOL-A, M2D, and BEATs were trained using AudioSet\cite{gemmeke2017audioset}, a large-scale dataset consisting of 2M samples of diverse sounds from YouTube audio tracks, and have been confirmed to be effective in audio tasks such as environmental sounds, speech, and music.
In contrast, OPERA-CT is trained exclusively on its curated respiratory sound dataset and has been validated using its OPERA benchmark, which includes various respiratory sound tasks.

\subsection{Evaluation Setup} \label{sec:exp-setup}
\begin{figure}[tbp]
  \centering
  \includegraphics[width=1.0\columnwidth]{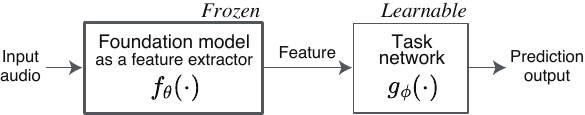}
  \vspace{-10pt}
  \caption{Network configuration for performance evaluation.}
  \label{fig:system}
  \vspace{-10pt}
\end{figure}

In the experiments, we set up a classifier network that uses a foundation model as a feature extractor, as illustrated in Fig. \ref{fig:system}. We trained the network on the training set from the evaluation task and then assessed its performance on the test set using the same procedures and metrics employed in previous SOTA studies.

We fixed the weights $\theta$ of the foundation model $f_\theta$, as shown `\textit{Frozen}' in Fig. \ref{fig:system}, and used its output features to train the weights $\phi$ of the task-specific network $g_\phi$. Since the features from the foundation model dictate the test results, we treat these results as the model's performance on the task.
For the task network, we used either the multilayer perceptron (MLP) or a more expressive 4-block transformer encoder.
As in the previous studies, we utilized the same hyperparameters for training on a task, such as the number of epochs, to facilitate comparison.
These hyperparameters and other details are described in the code provided online.

\subsection{Respiratory sound task: ICBHI2017} \label{sec:exp-ICBHI}
We use the ICBHI (International Conference on Biomedical and Health Informatics) 2017 Challenge (ICBHI2017)\cite{ICBHI2017} for a respiratory sound classification task and compare the performance of foundation models to the SOTA performance. ICBHI 2017 consists of 539 recordings, and the noise level of some recordings is high, reflecting real-life conditions. The task is to classify crackles, wheezes, a combination of them, or normal breathing.

We used the same experimental setup as in previous studies (e.g., official data split of the challenge and data augmentation method) \cite{moummad2022icbhi,M2D2024TASLP} and the codebase of \cite{M2D2024TASLP}.
We trained the classifier shown in Fig. \ref{fig:system} using the codebase.
The evaluation metrics are sensitivity and specificity for binary classification between normal breathing and other classes, as well as score, which is the average of sensitivity and specificity.
For the task network, we used the 4-block transformer encoder because the results with MLP were subpar in preliminary experiments.
As in \cite{M2D2024TASLP}, we calculated the statistics from five attempts to get the final results.

Table \ref{tab:exp:result-icbhi} shows that the results for BYOL-A, BEATs, and M2D in (iii) attain the baseline results shown in (i), demonstrating that these foundation models contribute to achieving baseline performance. However, they fall short of (ii) SOTA results.
Among these models, M2D achieves SOTA performance when fine-tuned, as shown in (ii), suggesting that fine-tuning is necessary for the foundation models to reach SOTA performance in this task.

Additionally, the ablation study of the task network with MLP, as shown in (iv), does not meet the baseline performance, indicating that the features of the foundation models are not effective for the task as they are, further supporting the need for fine-tuning.
In summary, we confirmed that three models, while not reaching SOTA, are useful in achieving baseline performance for ICBHI2017.

\begin{table}[tb!]
\caption{Respiratory sound classification (ICBHI2017) performance comparison.}
\label{tab:exp:result-icbhi}
\vspace{-5pt}
\centering
\resizebox{0.95\columnwidth}{!}{%
\begin{tabular}{llll}
\toprule
Model & Specificity & Sensitivity & Score (avg.)\\
\midrule
\multicolumn{4}{l}{\textit{(i) Baseline fine-tuning results}}  \\
Moummad\scriptsize{ et al. (CE) }\cite{moummad2022icbhi} & 70.09{\scriptsize $\pm$3.08} & 40.39{\scriptsize $\pm$2.97} & 55.24{\scriptsize $\pm$0.43} \\
Moummad\scriptsize{ et al. (SCL) }\cite{moummad2022icbhi} & 75.95{\scriptsize $\pm$2.31} & 39.15{\scriptsize $\pm$1.89} & 57.55{\scriptsize $\pm$0.81} \\

\multicolumn{4}{l}{\textit{(ii) SOTA fine-tuning results}}  \\
Bae\&Kim\scriptsize{ et al. (CE) }\cite{bae2023patchmix_icbhi} & 77.14 & 41.97 & 59.55 \\
Bae\&Kim\scriptsize{ et al. (MixCL) }\cite{bae2023patchmix_icbhi} & {81.66} & 43.07 & 62.37 \\
M2D (16$\times$16) \cite{M2D2024TASLP}   &  79.47{\scriptsize $\pm$2.10} &  42.93{\scriptsize $\pm$1.84} &       61.20{\scriptsize $\pm$0.46} \\
M2D (16$\times$4) \cite{M2D2024TASLP}   &{79.48{\scriptsize $\pm$1.47}}&{44.38{\scriptsize $\pm$1.55}}&{61.93{\scriptsize $\pm$0.25}}\\
M2D-X (16$\times$4) \cite{M2D2024TASLP}  &\textbf{81.51{\scriptsize $\pm$1.03}}& 45.08{\scriptsize $\pm$1.10} & {63.29{\scriptsize $\pm$0.24}}\\
BTS\cite{kim24BTS} & 81.40{\scriptsize $\pm$2.57} & {45.67{\scriptsize $\pm$2.66}}&\textbf{63.54{\scriptsize $\pm$0.80}}\\

\midrule
\multicolumn{4}{l}{\textit{(iii) Foundation models without fine-tuning (this study)}}  \\
AST\cite{gong2021ast} & 66.14{\scriptsize $\pm$2.47} & 43.72{\scriptsize $\pm$3.00} & 54.93{\scriptsize $\pm$0.78} \\
BYOL-A\cite{niizumi2023byol-a} & 71.49{\scriptsize $\pm$4.57} & 41.90{\scriptsize $\pm$4.83} & 56.70{\scriptsize $\pm$1.92} \\
BEATs$_{{iter3}}$ \cite{chen2022beats} & 72.05{\scriptsize $\pm$4.86} & 42.57{\scriptsize $\pm$5.96} & 57.31{\scriptsize $\pm$1.40} \\
M2D (16$\times$16) \cite{M2D2024TASLP} & 74.57{\scriptsize $\pm$3.72} & 41.82{\scriptsize $\pm$3.96} & 58.19{\scriptsize $\pm$1.36} \\
M2D (16$\times$4) \cite{M2D2024TASLP} & 72.20{\scriptsize $\pm$7.25} &\textbf{46.56{\scriptsize $\pm$5.61}}& 59.38{\scriptsize $\pm$0.94} \\
OPERA-CT\cite{OPERA} & 79.09{\scriptsize $\pm$9.51} & 18.86{\scriptsize $\pm$8.95} & 48.97{\scriptsize $\pm$1.21} \\

\midrule
\multicolumn{4}{l}{\textit{(iv) Ablation: MLP task network instead of transformer encoders}}  \\
M2D (16$\times$4, MLP) & 63.94{\scriptsize $\pm$1.92} & 45.81{\scriptsize $\pm$2.31} & 54.88{\scriptsize $\pm$0.65} \\

\bottomrule\\
\end{tabular}
}
\end{table}

\begin{table}[tb!]
\vspace{-10pt}
\caption{Respiratory sound classification (SPRS) performance comparison}
\label{tab:exp:result-sprs}
\vspace{-5pt}
\centering
\resizebox{0.95\columnwidth}{!}{%
\begin{tabular}{llll}
\toprule
Model & Specificity & Sensitivity & Score (avg.)\\
\midrule
\multicolumn{4}{l}{\textit{(i) Baseline fine-tuning results}}  \\
Moummad et al. (CE) \cite{moummad2022icbhi} & 76.89{\scriptsize $\pm$0.80} & 92.35{\scriptsize $\pm$1.10} &  84.62{\scriptsize $\pm$0.29} \\
Moummad et al. (SCL) \cite{moummad2022icbhi} & 80.69{\scriptsize $\pm$1.62} & 90.49{\scriptsize $\pm$1.27} & 85.59{\scriptsize $\pm$0.48} \\
\multicolumn{4}{l}{\textit{(ii) SOTA fine-tuning results}}  \\
M2D (16$\times$4) \cite{M2D2024TASLP} & 91.45{\scriptsize $\pm$0.82} & 86.10{\scriptsize $\pm$0.90} & 88.78{\scriptsize $\pm$0.43} \\
M2D-X (16$\times$4) \cite{M2D2024TASLP} &\textbf{93.13{\scriptsize $\pm$0.53}}&\textbf{86.42{\scriptsize $\pm$0.53}}&\textbf{89.77{\scriptsize $\pm$0.16}}\\

\midrule
\multicolumn{4}{l}{\textit{(iii) Foundation models without fine-tuning (this study)}}  \\
AST\cite{gong2021ast} & 90.73{\scriptsize $\pm$0.76} & 82.47{\scriptsize $\pm$1.71} & 86.60{\scriptsize $\pm$0.55} \\
BYOL-A\cite{niizumi2023byol-a} & 89.06{\scriptsize $\pm$0.84} & 77.94{\scriptsize $\pm$0.61} & 83.50{\scriptsize $\pm$0.40} \\
BEATs$_{{iter3}}$ \cite{chen2022beats} & 91.12{\scriptsize $\pm$2.01} & 85.86{\scriptsize $\pm$0.84} & 88.49{\scriptsize $\pm$0.82} \\
M2D (16$\times$16)\cite{M2D2024TASLP} & 90.50{\scriptsize $\pm$1.11} & 84.16{\scriptsize $\pm$2.14} & 87.33{\scriptsize $\pm$0.70} \\
M2D (16$\times$4)\cite{M2D2024TASLP} & 91.67{\scriptsize $\pm$0.86} & 84.47{\scriptsize $\pm$0.92} & 88.07{\scriptsize $\pm$0.53} \\
OPERA-CT\cite{OPERA} & 83.23{\scriptsize $\pm$4.72} & 60.00{\scriptsize $\pm$5.58} & 71.62{\scriptsize $\pm$0.84} \\

\midrule
\multicolumn{4}{l}{\textit{(iv) Ablation: MLP task network instead of transformer encoders}}  \\
M2D (16$\times$4, MLP) & 91.50{\scriptsize $\pm$2.55} & 61.44{\scriptsize $\pm$3.22} & 76.47{\scriptsize $\pm$1.13} \\

\bottomrule\\
\end{tabular}
}
\vspace{-10pt}
\end{table}

\subsection{Respiratory sound task: SPRSound (SPRS)} \label{sec:exp-SPRS}
Similar to ICBHI2017, SPRS is a respiratory sound classification task using a database of 2683 recordings. The experiment's codebase and settings are the same as those used in ICBHI2017. We used the same transformer encoder as in ICBHI2017 for the task network. Unlike ICBHI2017, the sounds for this task have less noise.

The results in Table \ref{tab:exp:result-sprs} differ from those of ICBHI2017, showing that some foundation models achieve SOTA performance without the need for fine-tuning. Foundation models in (iii), except OPERA-CT and BYOL-A, outperform the baseline in (i). Furthermore, BEATs and M2D (16$\times$4) are comparable to the SOTA performance in (ii).
However, as shown in the ablation (iv), the performance score degrades to approximately 76 when using the less expressive MLP as the task network, indicating that, as in ICBHI2017, the effectiveness of the foundation models as they are in this task is not sufficiently high.

The difference in performance trend from ICBHI2017 may stem from data, with SPRS being cleaner than ICBHI2017.
The ICBHI2017 samples were recorded using multiple recording devices and have a higher noise level, such as heart sounds layered over respiratory sounds.
For these noisy data, the audio features of the foundation models are likely to represent these noises as well as respiratory sounds. Therefore, fine-tuning would be more necessary with ICBHI2017 to reduce the impact of sounds other than respiratory ones.

\subsection{Heart sound task: CirCor} \label{sec:exp-CirCor}
For heart sound, we used the heart murmur classification task with the CirCor DigiScope dataset from the George B. Moody PhysioNet Challenge 2022\cite{GeorgeBMoodyPhysioNetChallenge2022,PhysioNet}.
Various approaches have been proposed for this task, and we have shown in \cite{Exploring} that fine-tuning foundation models can achieve SOTA performance.
We evaluated the models by training the network illustrated in Fig. \ref{fig:system} on the same codebase from \cite{Exploring} but using the models with fixed weights.
The data for this task have a high level of noise, reflecting real-case environments:
\textit{``Different noisy sources have been observed in our dataset, from the stethoscope rubbing noise to a crying or laughing sound in the background.''}

This task is a classification problem to predict the present/absent/unknown class of murmur, and the official metric, weighted accuracy (W.acc), gives higher values for the accuracy of present and unknown classes.
$\text{W.acc}=(5c_p+3c_u+c_a)/(5t_p+3t_u+t_a)$, where $c_i$ and $t_i$ for $i\in\{p\text{(resent)},u\text{(nknown)},a\text{(bsent)}\}$ are the number of correct results and true labels, respectively.
Unweighted average recall (UAR) is another metric that emphasizes the importance of recalling all classes evenly.
The available data comprised 3163 recordings from 963 patients.
We followed \cite{Exploring} for the experimental setup, including data splits, training settings, and calculation of statistical results.

The results in Table \ref{tab:exp:result-circor} show that none of the models reach the SOTA performance (over 0.8) for weighted accuracy (W.acc), although the M2D model demonstrates effectiveness with results close to SOTA (over 0.7) for unweighted accuracy (UAR). 
On the other hand, all models show lower recall for the presence of murmur than the fine-tuning results, while they exhibit higher recall for unknown cases.
In particular, when using BEATs, most samples are predicted as unknown, making it difficult for this task.
In summary, similar to the fine-tuning results in \cite{Exploring}, M2D proved effective for this task.

\begin{table}[tb!]
\caption{Heart murmur recognition (CirCor) task performance comparison.}
\label{tab:exp:result-circor}
\vspace{-5pt}
\centering
\resizebox{1.0\columnwidth}{!}{%
\begin{tabular}{llllll}
\toprule
       &        &        &  \multicolumn{3}{c}{Recall} \\
\cmidrule(lr){4-6}  
Model  &   W.acc &    UAR &  Present &  Unknown &  Absent \\
\midrule
\multicolumn{5}{l}{\textit{(i) SOTA fine-tuning results (no foundation models used)}}  \\
Panah et al. \cite{Panah2023Exploring} & 0.80 & 0.70 & 0.86 & 0.41 & 0.83 \\
CUED\cite{CUED_acoustics2022} & 0.80 & 0.68 &\textbf{0.93}& 0.34 & 0.78 \\

\multicolumn{5}{l}{\textit{(ii) SOTA fine-tuning results from \cite{Exploring} using foundation models}}  \\
AST\cite{gong2021ast} & 0.654{\scriptsize $\pm$0.042} & 0.672{\scriptsize $\pm$0.028} & 0.744{\scriptsize $\pm$0.050} & {0.769{\scriptsize $\pm$0.069}}& 0.505{\scriptsize $\pm$0.101} \\
BYOL-A\cite{niizumi2023byol-a} & 0.556{\scriptsize $\pm$0.043} & 0.556{\scriptsize $\pm$0.025} & 0.590{\scriptsize $\pm$0.112} & 0.573{\scriptsize $\pm$0.112} & 0.507{\scriptsize $\pm$0.098} \\
BEATs$_{{iter3}}$\cite{chen2022beats}  &{0.809{\scriptsize $\pm$0.015}}& 0.673{\scriptsize $\pm$0.018} &{0.893{\scriptsize $\pm$0.044}}& 0.267{\scriptsize $\pm$0.059} & 0.860{\scriptsize $\pm$0.025} \\

M2D\cite{M2D2024TASLP} &\textbf{0.832{\scriptsize $\pm$0.009}}& \textbf{0.713{\scriptsize $\pm$0.027}}& 0.911{\scriptsize $\pm$0.012} & 0.361{\scriptsize $\pm$0.090} & {0.868{\scriptsize $\pm$0.011}}\\

\midrule
\multicolumn{5}{l}{\textit{(iii) Frozen foundation model results (this study, no fine-tuning)}}  \\

AST\cite{gong2021ast} & 0.604{\scriptsize $\pm$0.024} & 0.631{\scriptsize $\pm$0.020} & 0.776{\scriptsize $\pm$0.037} & 0.788{\scriptsize $\pm$0.037} & 0.327{\scriptsize $\pm$0.022} \\
BYOL-A\cite{niizumi2023byol-a} & 0.618{\scriptsize $\pm$0.019} & 0.566{\scriptsize $\pm$0.029} & 0.618{\scriptsize $\pm$0.047} & 0.396{\scriptsize $\pm$0.102} & 0.683{\scriptsize $\pm$0.057} \\
BEATs$_{{iter3}}$ \cite{chen2022beats} & 0.381{\scriptsize $\pm$0.126} & 0.493{\scriptsize $\pm$0.070} & 0.431{\scriptsize $\pm$0.177} & \textbf{0.878{\scriptsize $\pm$0.103}} & 0.170{\scriptsize $\pm$0.139} \\

M2D\cite{M2D2024TASLP} & 0.737{\scriptsize $\pm$0.023} & 0.693{\scriptsize $\pm$0.024} & 0.676{\scriptsize $\pm$0.044} & 0.525{\scriptsize $\pm$0.051} &\textbf{0.879{\scriptsize $\pm$0.018}}\\

\bottomrule
\end{tabular}
}
\end{table}

\subsection{Heart sound task: BMD-HS} \label{sec:exp-BMD-HS}
As another heart sound task, we utilized the BUET Multidisease Heart Sound (BMD-HS) dataset\cite{BMD-HS}.
In contrast to CirCor, this task features data with less noise:
\textit{``PCGs were re-recorded if any ambiguity or noise was detected in the signal.''}

Although the class definition contains multiple diseases, we treated it as a binary classification to predict the presence or absence of a murmur, allowing us to obtain results that could be compared to the baseline. The metrics used were accuracy, sensitivity, specificity, and F1 score.
The dataset consists of 864 recordings from 108 subjects, and we created three splits of training and test sets with stratification under different random seeds and obtained statistical results using the results of these data splits. Specifically, we assigned 576 recordings to the training set and 288 to the test set and tested five times for each split to average 15 results as the final statistical results.

The experimental results in Table \ref{tab:exp:result-bmdhs} demonstrate that ASTs, BEATs, and M2D outperform fine-tuning baseline, demonstrating their effectiveness.
Similar to the tasks involving respiratory sounds, we confirmed that the foundation models are effective with clean data (BMD-HS) compared to noisy data (CirCor) for heart sounds.

\begin{table}[tb!]
\caption{Heart sound classification (BMD-HS) task performance comparison}
\label{tab:exp:result-bmdhs}
\vspace{-5pt}
\centering
\resizebox{1.0\columnwidth}{!}{%
\begin{tabular}{llllll}
\toprule
Model  &   Accuracy & Sensitivity & Specificity & F1 Score \\
\midrule
\multicolumn{5}{l}{\textit{(i) Baseline fine-tuning results (no foundation models used)}}  \\
Baseline\cite{BMD-HS} & 0.80 & 0.88 & 0.75 & 0.80 \\

\midrule
\multicolumn{5}{l}{\textit{(ii) Frozen foundation model results (this study)}}  \\
AST\cite{gong2021ast} & 0.924{\scriptsize $\pm$0.008} & 0.868{\scriptsize $\pm$0.024} & 0.937{\scriptsize $\pm$0.008} & 0.816{\scriptsize $\pm$0.018} \\
BYOL-A\cite{niizumi2023byol-a} & 0.870{\scriptsize $\pm$0.013} & 0.592{\scriptsize $\pm$0.113} & 0.937{\scriptsize $\pm$0.022} & 0.620{\scriptsize $\pm$0.071} \\
BEATs$_{{iter3}}$\cite{chen2022beats} &\textbf{0.952{\scriptsize $\pm$0.006}}& 0.879{\scriptsize $\pm$0.036} &\textbf{0.969{\scriptsize $\pm$0.005}}&\textbf{0.875{\scriptsize $\pm$0.018}}\\
M2D\cite{M2D2024TASLP} & 0.933{\scriptsize $\pm$0.008} &\textbf{0.890{\scriptsize $\pm$0.027}}& 0.943{\scriptsize $\pm$0.008} & 0.838{\scriptsize $\pm$0.020} \\

\bottomrule
\end{tabular}
}
\vspace{-10pt}
\end{table}

\section{Discussion}
\textbf{The foundation models designed for generic audio tasks outperformed the specialized model.}
The respiratory sound model OPERA-CT underperforms other general audio models in the respiratory sound tasks.
For example, OPERA-CT, shows a score of 48.97 in Table \ref{tab:exp:result-icbhi} in ICBHI2017, which is approximately 10 points less than the score of 59.38 for M2D (16$\times$4).
A possible reason for the discrepancy could be the difference in the training data. General audio models, such as M2D, are trained on AudioSet, which features 2M diverse and large-scale samples. In contrast, OPERA-CT was trained on only 140K monotonic respiratory sounds.

\vspace{0.05cm}\textbf{The features of the audio foundation models can be more effective in tasks with clean data for both respiratory and heart sounds.}
The experiments confirmed the effectiveness in SPRS and BMD-HS tasks, where data are clean compared to ICBHI2017 and CirCor.
If the data are clean, the general-purpose features are more likely to be useful in solving a problem; in contrast, fine-tuning is considered necessary to make a model avoid the noise and produce output features that focus on the target sound of the task if the data is noisy.
This observation suggests that preprocessing or cleansing sounds before extracting their features utilizing a denoising, signal decomposition\cite{PPGDecomposition, EEGSubbands}, or target sound extraction technique such as \cite{SoundBeam} and \cite{SoundBeamM2D} would be advantageous for making the feature of foundation models more effective.

\vspace{0.05cm}\textbf{Effective foundation models are pre-trained on the latest self-supervised learning methods.}
Among these models, M2D demonstrated consistent effectiveness across all tasks, while BEATs showed the best effectiveness in datasets with less noise, such as SPRS and BMD-HS. The experimental results indicate that analyzing respiratory and heart sounds requires the latest audio representation learning paradigm using a large-scale and diverse general audio dataset to be effective.

\section{Conclusion}
This paper evaluated the effectiveness of fixed-weight audio foundation models in respiratory and heart sound analysis by comparing their results with previous SOTA results in two respiratory tasks, ICBHI2017 and SPRS, and two heart sound tasks, CirCor and BMD-HS.
While the foundation models did not achieve SOTA performance in ICBHI2017 and CirCor, where the data are noisy, they contributed to achieving SOTA performance in SPRS and BMD-HS, where the data are less noisy.
Furthermore, the respiratory sound task experiments showed that the general audio models outperform a model trained only on respiratory sounds.
Future research may benefit from incorporating denoising, signal decomposition, and sound source separation techniques, along with fine-tuning approaches and a more in-depth analysis of errors such as misclassifications.

\addtolength{\textheight}{-12cm}   




\bibliographystyle{IEEEtran}
\bibliography{refs}

\end{document}